\documentclass[10pt]{article}

\usepackage{amsmath}
\usepackage{amssymb}

\usepackage{graphicx}

\usepackage{cite}

\usepackage{color}

\usepackage[labelfont=bf,labelsep=period,justification=raggedright]{caption}

\makeatletter
\renewcommand{\@biblabel}[1]{\quad#1.}
\makeatother

\date{}

\begin{document}

\begin{flushleft}
{\Large
\textbf{Characterizing and Modeling the Dynamics of Activity and Popularity}
}
\\
Peng Zhang$^{1}$,
Menghui Li$^{2,3\ast}$,
Liang Gao$^{4}$,
Ying Fan$^{2}$,
Zengru Di$^{2}$
\\
\bf{1} School of Science, Beijing University of Posts and
Telecommunications, Beijing 100876, P. R. China
\\
\bf{2} School of Systems Science, Beijing Normal University,
Beijing 100875, P. R. China
\\
\bf{3} Beijing Institute of Science and Technology Intelligence, Beijing, 100048, P. R. China
\\
\bf{4} Systems Science Institute and MOE Key Laboratory for Urban
Transportation Complex Systems Theory and Technology, Beijing Jiaotong
University, Beijing 100044, P. R.  China
\\
$\ast$ E-mail: limenghui76@gmail.com
\end{flushleft}

\section*{Abstract}
Social media, regarded as two-layer networks consisting of users and items, turn out to be the most important channels for access to massive information in the era of Web 2.0. The dynamics of human activity and item popularity is a
crucial issue in social media networks. In this paper, by analyzing the growth of user activity and item popularity in four empirical social media networks, i.e., Amazon, Flickr, Delicious and Wikipedia, it is found that cross links between users and items are more likely to be created by active users and to be acquired by popular items, where user activity and item popularity are measured by the number of cross links associated with users and items. This indicates that users generally trace popular items, overall. However, it is found that the inactive users more severely trace popular items than the active users. Inspired by empirical analysis, we propose an evolving model for such networks, in which the evolution is driven only by two-step random walk. Numerical experiments verified that the model can qualitatively reproduce the distributions of user activity and item popularity observed in empirical networks. These results might shed light on the understandings of micro dynamics of activity and popularity in social media networks.

\section*{Introduction}
In recent years, social media networks, vital platforms for sharing contents with others in the era of Web 2.0, such as YouTube, Facebook,
Delicious, Amazon, Flickr and Wikipedia, to name just a few, have experienced explosive growth \cite{socialmedia,socialmedia1}. These
systems record the fingerprints of every user's activity and every
item's popularity, providing ``a wealth of data" to study the dynamics
of human activity and item popularity at the global system scale.
In particular, it is found that the probability distributions of the activity degree of users, e.g., editing in Wikipedia \cite{activity}, voting
in News2 \cite{activity} as well as favorite marking in Flickr
\cite{Flickr}, and the popularity degree of items, e.g., the number of fans a photo has in Flickr \cite{Flickr}, follow a power law.
The power law distributions are explained by the
rich-get-richer mechanism \cite{rich,rich1}, which is also called
preferential attachment in the field of complex networks
\cite{CN:REV2002,CN:REV2003,CN:REV2006}. However, how these two distributions arise simultaneously due to human activity has yet to be determined.

The activity dynamics \cite{activity,humandynamics,socialdynamics,Editorial,roleofactivity,humanactivity} and popularity dynamics \cite{Ratkiewicz2010,popularitys,popularitys1,popularitys2,popularitys3,citationdynamics,meme,Content,socialinfluence1}
 have been investigated in the literatures, respectively. However, human
activity and item popularity, two perspectives of the cross links between users and
items, are interdependent; therefore, we can not study
the dynamics of one aspect alone. In addition, individuals are
always embedded in a social network. It is widely believed that
information can spread quickly along social links using user-to-user exchanges, also known as "word-of-mouth" exchanges; moreover, the users' behaviors are strongly influenced by their neighbors \cite{socialinfluence1,socialinfluence,socialinfluence2,healthbehavior,socialinfluence3}. In particular, the social degree and the activity degree depend on each other \cite{activity}. Hence, it is considered worthwhile studying social networks to obtain deeper insights into the dynamics of human activity and item popularity. Until now, there has been no clear picture as to how online human activity and item popularity coevolve, so it is crucial to
investigate the evolution of empirical human activity and item
popularity as well as the theoretical model to obtain a better
understanding of the possible generic laws governing the formation of
activity distribution and popularity distribution.

In this paper, we first characterize the evolution of human
activity and item popularity in the Amazon, Flickr, Delicious and
Wikipedia networks. It is found that in such social media networks, both
relative probabilities of users creating cross links and items
acquiring cross links are proportional to the degree of activity and
degree of popularity, respectively. In particular, the inactive users
are more likely to trace popular items than the active users. Based on
empirical observations, we then propose an evolving model based on
two-step random walk. Finally, we justify the validity of our model by
comparing the results of model with that of empirical networks.
This work could shed light on the understanding of evolution of user activity and item popularity in social media networks, and it also could be helpful in certain applications, such
as designing efficient strategies for virtual marketing and network
marketing, etc.

\begin{figure}[!ht]
\begin{center}
\includegraphics[width=3in]{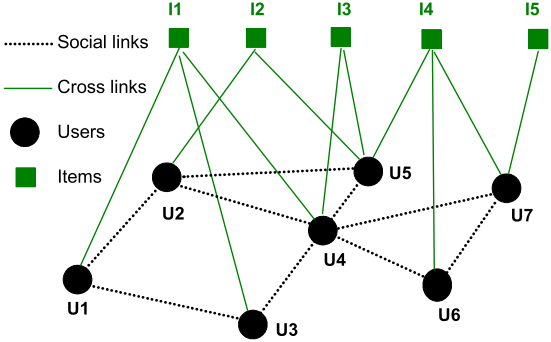}
\end{center}
\caption{{\bf Schematic plot of the social media networks.}
For these two types of links, we define three types of degrees.
For example, U4 in the network has social degree ($k_{s}=5$), and
the activity degree ($k_a=2$); I4 has the popularity degree ($k_p=3$). Please note that there is no social links in some cases, such as Wikipedia.}\label{fig1}
\end{figure}

\section*{Materials and Methods}
\textbf{Data description and notations.} The Delicious data set
was downloaded from http://data.dai-la bor.de/corpus/delicious/, and
consists of 132,500,391 bookmarks, 50,221,626 URLs (books), and
947,835 users between September, 2003 and December 31, 2007  \cite{delicious}.
The Amazon user-movie rating data set was obtained from Stanford Large Network Dataset Collection (http://snap.stanford.edu/data/web-A mazon.html) \cite{amazon}. The data
consists of 7,911,684 ratings, 267,320 movies and 759,899 users
between August 1997 and October 2012. The Flickr data set was
collected by daily crawling Flickr over 2.5 million users from Nov
2, 2006 to Dec 3, 2006, and again daily from February 3, 2007 to May 18,
2007 (http://socialnetworks.mpi-sws.org/datasets.html)\cite{Flickr}. Here, we only considered the users who had at least one favorite photo. With this constraint, there are 497,937 users, 11,232,836 photos and 34,734,221 favorite-markings in the data. The Wikipedia (English) data set was download from
http://konect.uni- koblenz.de/networks/edit-enwiki. The data set consists
of 21,416,395 articles written collaboratively by 3,819,691
volunteers around the world before September, 2010. The four
datasets consist of individuals and items, such as movies in Amazon,
URLs (books) in Delicious, photos in Flickr, and articles in Wikipedia.
Moreover, users are able to show interest in these items using the network feature of rating in Amazon, bookmarking in Delicious, favorite-marking in
Flickr, and editing in Wikipedia. Therefore, these systems are
topologically equivalent. For analysis purposes, the user-item data
can be mapped into a two-layer network, as shown schematically in
Fig. \ref{fig1}. This network has two types of nodes: $M$ users and $N$
items totally. In principle, the individuals are embedded in a
social network. For example, Flickr and Delicious allow users to make friends. Therefore, there should be two types of links: the cross links between users and items as well as the social links among users.

Mathematically, the topology shown in Fig. \ref{fig1} can be
characterized by two matrices. $S$, an $M \times M$ adjacency
matrix, represents the social links among users, with element
$S_{ij} = 1$ if user $i$ declares user $j$ as his friend, otherwise
$S_{ij} = 0$. Similarly, $C$, an $M \times N$ adjacency matrix, characterizes
the cross links, with element $C_{i\lambda}=1$ if user $i$ is
interested in the item $\lambda$, otherwise $C_{i\lambda}=0$. To be specific, we defined the following types of degrees to characterize the
multi-relational connections. Two degrees are related to the cross
links: (1) the activity degree: $k_a(i)=\sum_{\lambda}C_{i\lambda}$,
i.e., the number of items interested by user $i$; (2) the
popularity degree $k_p(\lambda)=\sum_{i}C_{i\lambda}$, i.e., the
number of users who are interested in the item $\lambda$, which
reasonably represents the popular extent in the network; and (3) the social degree $k_s=\sum_{j}S_{ij}$, i.e., the number of friends for a given user.
Note, $k_a$ and $k_p$ are two different perspectives of the cross links
connecting users and items.

\textbf{Measuring preferential attachment.} Here, we explain the
method for measuring the phenomenon of preferential attachment on
temporal data \cite{preferential,preferential1}. The basic idea is to investigate whether new links
are likely to attach to nodes with larger degree (size). We
calculate the empirical value of the relative probability $\Pi(k_T)$
that a new cross link formed within a short period $\Delta t$
connects to a user (item), which has a degree of $k_T$ at the time
$t_0$, as follows,
\begin{equation}
\Pi(k_T)=\frac{\frac{A(k_T)}{C(k_T)}}{\sum_{k'_T}\frac{A(k'_T)}{C(k'_T)}}.\label{limh}
\end{equation}
Here, $k_T$ is the degree at time $t_0$.
$A(k_T)=\sum^{k_T(t)=k}_{i,\lambda}C_{i\lambda}$ is the number of
nodes with exact degree $k_T$ at $t_0$, but creating (acquiring) new
cross links within next small interval $\Delta t$ (e.g., one day in
this article). $C(k_T)$ is the number of users (items) with degree
$k_T$ at $t_0$. The preferential attachment hypothesis states that the rate
$\Pi(k_T)$ with which a node with $k_T$ links acquires new links is
a monotonically increasing function of $k_T$ \cite{BAmodel}, namely
$\Pi(k_T)\sim k_T ^ \alpha$. To obtain a smooth curve from noisy
data, we take the cumulative function form instead of $\Pi(k_T)$:
\begin{equation}
\kappa\left(k_T\right) = \int_0^k \Pi(k'_T)\,dk^{'}_T\sim
k_T^{\alpha + 1}.
\end{equation}
In our measurement, $k_T$ can be either degree of activity $k_a$ or
degree of popularity $k_p$. This method has been successfully used to
verify the preferential attachment mechanism of BA model
\cite{BAmodel} in empirical evolving networks
\cite{preferential,preferential1,tracehot,microdynamics} and
theoretical models \cite{measurePA}.

\begin{table}[!ht]
\caption{Basic statistics of the data sets constructed for our study. Showing the number of users $M$, the number of items $N$, the number
of cross links $E$, the average degree of activity $\langle k_a\rangle$
and the average degree of popularity $\langle k_p\rangle$}
\begin{center}
\begin{tabular}{|c|c|c|c|c|c|}
\hline
 & ~~~~$M$~~~~& ~~~~$N$~~~~&~~~~$E$~~~~&~~~~$\langle k_a\rangle$~~~~&~~~~$\langle k_p\rangle$~~~~\\
\hline
Amazon & 759,899& 267,320&7,911,684&10.41&29.59\\
\hline
Flickr  & 497,937 & 11,232,836 &34,734,221&69.80&3.09\\
\hline
Wikipedia &3,819,691 & 21,416,395 &122,075,170&31.96&5.70\\
\hline
Delicious&947,835&50,221,626& 132,500,391& 139.80&2.64\\
\hline
\end{tabular}\label{dataset}
\end{center}
\end{table}

\textbf{Measuring relative contributions ratio.} To measure the
relative contribution ratio within a small interval $\Delta t$
(e.g., one day in this article), we extend the method proposed in
the reference \cite{tracehot} as follows. Absolute contribution from
the users with degree $k_a$ is measured simply by a percentage of
new cross links created by these users within a short period
$[t_0,t_0+\Delta t]$ out of the total number of cross links at time
$t_0$,
\begin{equation}
r^{k_a} = \frac{\sum_{k_p} H^{k_a}_{k_p}}{\sum_{k_p} H_{k_p}},
\end{equation}
where $H^{k_a}_{k_p}=\sum^{k_i=k_a}_{k_{\lambda}=k_p}C_{i\lambda} =A^{k_a}(k_p)$
is the number of new links, which are created by the users with
degree $k_a$ and attached to the items with degree $k_p$, within the
period $\Delta t$ (e.g, one day in this paper).
$H_{k_p}=\sum^{k_{\lambda}=k_p}_{\lambda}k_{\lambda}= C(k_p)k_p$ is the number
of cross links attached to the items with degree $k_p$ at time
$t_0$. In order to observe the  differences of users' activity with
different degree $k_a$, we present a more detailed breakdown of this
absolute contribution by calculating the percentage of new cross links, which are from the users
with degree $k_a$ and attach to the items with degree $k_p$, out of all links
attached to these items at time $t_0$, and then normalized
by the absolute contribution of these users, namely relative
contribution ratio,
\begin{equation}
R^{k_a}(k_p)= \frac{H^{k_a}_{k_p}}{H_{k_p}}\frac{1}{r^{k_a}}. \label{RCR}
\end{equation}
In a sense, it describes how often the users with degree $k_a$ are
pursuing popular items.  In principle, the Eq. (\ref{limh}) is related to the Eq. (\ref{RCR}) for the users with degree $k_a$ as follows
\begin{equation}
\Pi^{k_a}(k_p) \sim \frac{A^{k_a}(k_p)}{C(k_p)} = r^{k_a}k_p R^{k_a}(k_p).\label{rewrite}
\end{equation}

\section*{Results}
\subsection*{Empirical Analysis to Temporal Data }
As shown in Fig. \ref{fig1}, Amazon, Flickr, Delicious and Wikipedia
are typical social media networks consisting of users and items such
as movies, photos, books, articles, etc (see \emph{Materials and Methods} for data
description and notations). Social media networks are more
complicated than  the networks with one type of links in previously studies \cite{CN:REV2002,CN:REV2003,CN:REV2006}, including single node networks and bipartite networks, due to their multiplex nodes and
multi-relations. Basic statistical properties for each data set are
shown in Table \ref{dataset}. The degree of activity
and degree of popularity follow an approximately power law distribution \cite{Flickr,amazon,delicious}. In
particular, the social degree follows a power law distribution in Flickr
\cite{Flickr}. In the
following, we report the main findings of our empirical analysis of
the Amazon, Flickr, Delicious and Wikipedia networks. We pay particular attention to the evolution of activity degree and popularity degree in these four networks.

\begin{figure}[!ht]
\begin{center}
\includegraphics[width=5.5in]{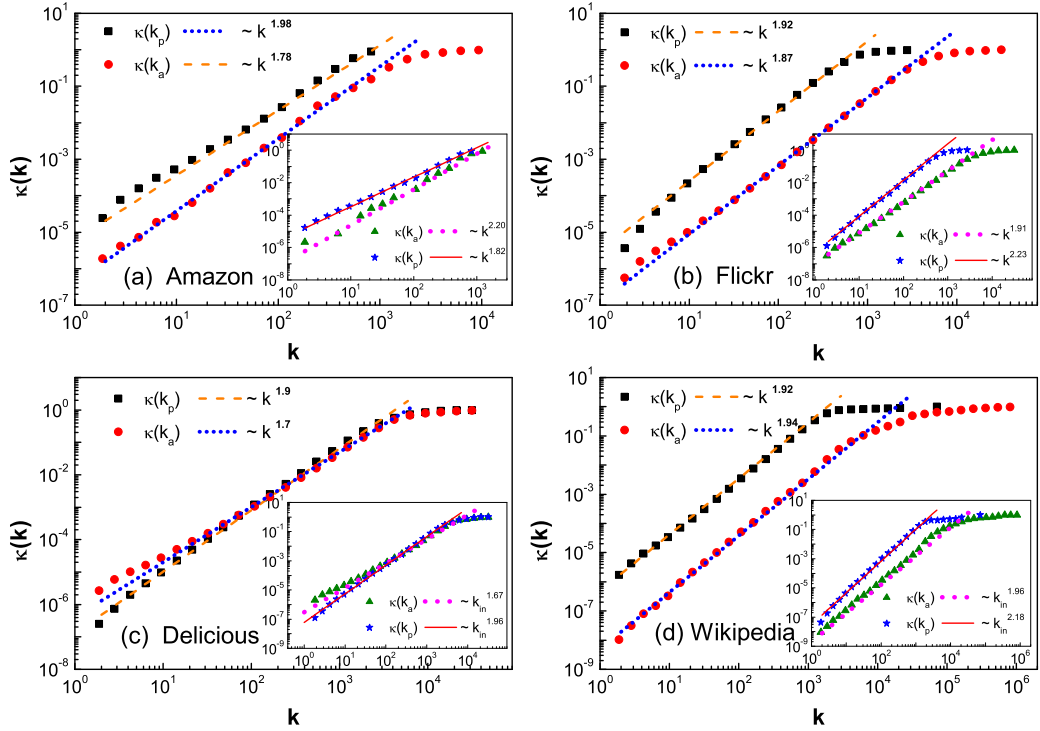}
\end{center}
\caption{ {\bf Influence of current degree on the formation of new cross links.} The cumulative functions of relative probability $\kappa (k_a)$ and $\kappa (k_p)$, characterizing
the influence of current activity degree and popularity degree on the formation of cross links between existing users and existing items respectively, {where the time interval $\Delta t$ is one day. } (a) Amazon for the snapshot of 01/06/2012, (b) Flickr for the snapshot of 01/03/2007, (c) Delicious for the snapshot of 01/12/2007 and  (d) Wikipedia for the snapshot of 01/09/2010. In the insets, $\kappa (k_a)$ are the relative probability of existing users creating cross links to new items, and $\kappa (k_p)$ are the relative probability of existing items
acquiring cross links from new users. Through this paper, the results of empirical analysis are for the same snapshot, and the similar patterns are also observed in other snapshots of the data. The straight lines are guide
to the eye. }\label{Figure-2}
\end{figure}

Like many other complex networks, the growths of these four networks
involve two major factors: adding new nodes and generating new
links. Here we pay particular attention to the formation of new
links during the evolution of networks, because this is the
central process in which users can exchange information with each
other. In the following, we focus on how the existing states of
users and items affect the formation of new links and what encompasses the differences between various users' interests.

First, we examine the phenomenon of preferential creation on the existing users and
preferential attachment on the existing items in
these four data sets. To this end, we employ a numerical method,
proposed to test preferential attachment (see \emph{Materials and Methods} for more
details), to investigate how the generation of new cross links depends on the
existing degrees in the temporal data sets.

\begin{table}[!ht]
\caption{Exponents $\alpha_a$ and $\alpha_p$ for preferential creation and preferential attachment. Exponents $\alpha_a$ and $\alpha_p$ as in $\kappa (x) \sim
x^{\alpha+1}$, which characterize the influence of current degree of activity
and degree of popularity on the formation of cross links between
existing users and existing items as well as the formation of cross
links between existing users (items) and new items (user) [in the
brackets] The results are averaged over 10 randomly selected snapshots, where the exponents are determined by least-square fitting, and $R^2>0.99$ generally.}
\begin{center}
\begin{tabular}{|c|c|c|c|c|}
\hline
 & ~~~~Amazon~~~~& ~~~~Flickr~~~~&~~~~Delicious~~~~&~~~~Wikipedia~~~~\\
\hline
$\alpha_a$ &0.78 [1.20]& 0.87 [0.91]&0.7 [0.67]&0.94 [0.96]\\
\hline
$\alpha_p$  & 0.98 [0.82] & 0.92 [1.23] &0.9 [0.96]&0.92 [1.18]\\
\hline
\end{tabular}
\end{center}\label{alpha}
\end{table}

Figure \ref{Figure-2} shows the cumulative function $\kappa (k)$
with respect to the degree of activity and degree of popularity. We see that the relative cumulative probability $\kappa(k_a)$ ($\kappa (k_p)$) for users (items) to create (acquire)
cross links is proportional to the existing degree of activity (popularity). In particular, the cumulative functions $\kappa$ approximately follow a straight line on the log-log scale,
indicating that the relative cumulative probability of generating new degrees
satisfies a power law with respect to the existing degrees, which
can be characterized by the positive exponent $\alpha$ where $\kappa(x) \sim x^{\alpha+1}$ with $x$ denoting the degree. In Table \ref{alpha}, we list the characteristic exponents $\alpha_a$
and $\alpha_p$ determined by least-square fitting the $\kappa$ functions for small $k$ as the curves deviate from the straight line for large $k$ due to low statistics. The positive
exponents $\alpha_a$ and $\alpha_p$ indicate that the active users
(with a higher degree of activity) have greater chance to create new cross links than the inactive users (with a lower degree of activity), while the popular items (with a higher degree of popularity ) have greater chance to attract new cross links.

\begin{figure}[!ht]
\begin{center}
\includegraphics[width=5.5in]{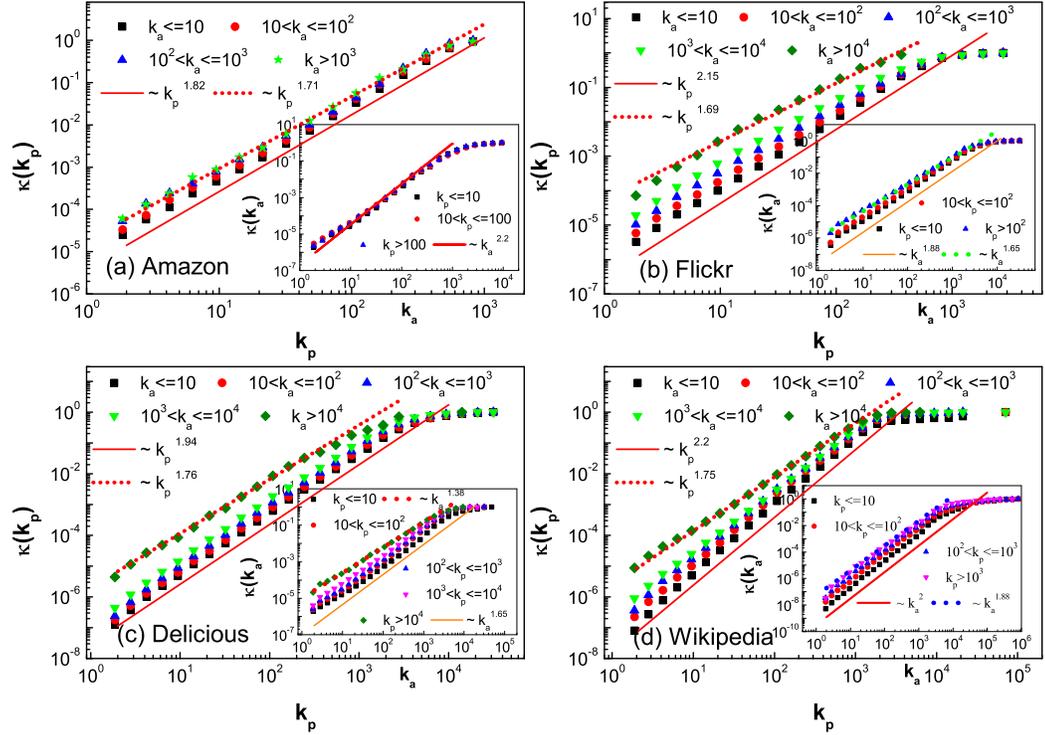}
\end{center}
\caption{ {\bf Influence of degree on the intensity of tracing popular item and attracting attentions.} The cumulative function of relative
probability $\kappa(k_p)$ for different groups of users and
$\kappa(k_a)$ for different groups of items (in the insets), in
Amazon(a), Flickr (b), Delicious (c) and Wikipedia (d). The users and items are classified into different groups according to
the degree of activity and the degree of popularity. }\label{fig3}
\end{figure}

As these four systems expand rapidly,
we then investigate the formation of cross links between new
users (items) and existing items (users). In the insets of Fig. \ref{Figure-2}, $\kappa(k_a)$ characterizes the relative probability that the existing users are interested in new items with
respect to the users' degree of activity, whereas $\kappa(k_p)$ characterizes the relative probability that the existing items attract the attentions of new users with respect to the items' degree of popularity. Interestingly, as seen in the insets of Fig. \ref{Figure-2}, these cumulative functions  also follow a power law. The positive exponents $\alpha_a$ and $\alpha_p$ indicate that the newly created items are more likely to attract the attentions of active users, while the new users are more likely to be interested in popular items. The above results suggest that the users are likely to trace popular items overall, and that the active users are more likely to create new cross links than the inactive users.

\begin{figure}[!ht]
\begin{center}
\includegraphics[width=5.5in]{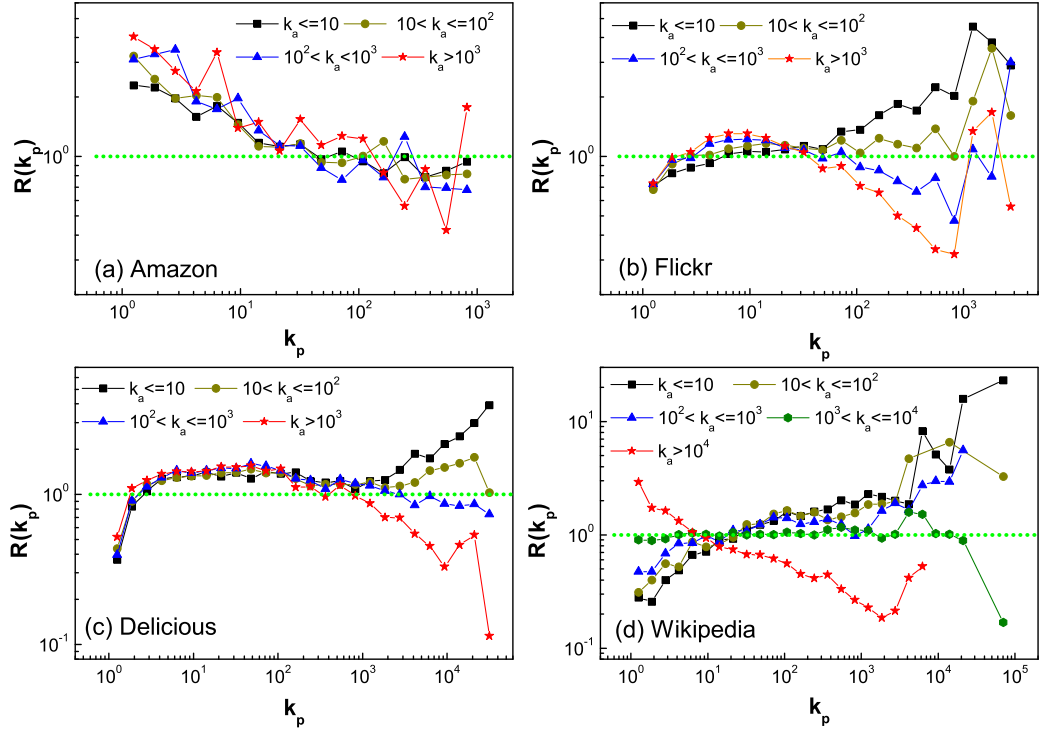}
\end{center}
\caption{{\bf Results of the relative contribution ratios $R(k_p)$ for users with different activity degree.} (a) Amazon, (b) Flickr, (c) Delicious and (d) Wikipedia. The users
are classified into following types according to different degrees of activity, i.e., $k_a \leq10$, $10<k_a\leq 100$, $100< k_a \leq 1000$, $1000<k_a \leq
10000$ and $k_a>10000$. The values of $R(k_p)$ 	
fluctuate significantly for large $k_p$ due to low statistics.}\label{fig4}
\end{figure}

What is the influence of activity (popularity) degree on the intensity of users tracing popular items (items attracting attentions of users)? To attack this problem, we classify the users and items into different groups according to their activity degree and popularity degree. Then, we investigate the cumulative
functions of relative probability $\kappa(k_p)$ and $\kappa(k_a)$ for different group of users and items, respectively.  As seen from Fig. \ref{fig3}, the slops of inactive users (with smaller $k_a$) tracking items look qualitatively larger than those of active users. For instance, in Wikipedia, the slope is 2.2 for $k_a \leq 10$, while it is 1.75 for $k_a>10000$.  This indicates that the inactive users more severely
trace popular items than the active users.  Moreover, as seen in the insets of Fig. \ref{fig3}, the slops of unpopular items (with smaller $k_p$)
attracting users look slightly larger than those of popular items, indicating
that the unpopular items attract a greater interest amongst active users. Please note that the differences between different groups of users or items are smaller for Amazon than for the other three networks. This may be due to
the different spreading modes of items such as movies. For
instance, a popular movie is similar to the well-known global information in the Amazon user-movie network, but there is no such counterpart in the Flickr user-photo, Delicious user-book and
Wikipedia author-article networks.

To provide an additional evidence for the different intensity of users
tracing popular items, we also calculate the relative
contribution $R^{k_a}(k_p)$ of users with
activity degree $k_a$, who create some new cross links to items with popularity degree
$k_p$ within one day (see \emph{Materials and Methods} for the detail). Ideally, if the intensities of users tracing popular items are identical, the relative contribution ratios $R(k_p)$ should be always equal to 1 for all group of users. As seen in Fig. \ref{fig4}, the relative contribution of active users (with larger $k_a$) to unpopular items (with smaller $k_p$) is larger than 1 but is smaller than 1 for popular items (with larger $k_p$), indicating that the active users make higher contributions to unpopular items than average but lower to popular items than average. Meanwhile, the inactive users exhibit the opposite behavior with the exception of Amazon. Especially in Flickr and Wikipedia, it is obviously found that $R(k_p)$ increases for the most inactive users, while $R(k_p)$  decreases for the most active users with respect to popularity degree. Based on Eq. (\ref{rewrite}), the cumulative function $\kappa(k_p)$ for the inactive users will increase more faster than that for the active users, indicating that the slop for inactive users is larger than that for active users (as shown in Fig. \ref{fig3}). Furthermore, it is found
that medium active users within Wikipedia, e.g., $1000<k_a \leq 10000$, make almost equivalent contributions to articles having different degree of popularity (as shown in Fig.\ref{fig4} (d)), indicating that they may not
care about the article's popularity when they edit them. These results also
indicate that the inactive users are more likely to trace popular
items than the active users, in agreement with previous observations.

\subsection*{Modeling}
To further understand the mechanisms governing the evolution of real networks,
we attempt to set up a theoretical model for
user-item networks. Our primary goal is to qualitatively reproduce
the human activity and the item popularity observed in the four
empirical networks previously mentioned. In the above numerical analysis, the rich-get-richer phenomenon has been observed in the growth of the user's degree of activity and the item's degree of popularity, i.e., the active users (the popular items) have a higher probability of creating (acquiring) new cross links.The mechanism of preferential attachment has successfully explained the rich-get-richer phenomenon in previous works \cite{CN:REV2002,CN:REV2003,CN:REV2006}, but it implicitly requires global information, e.g. degree of all nodes. However, it is impossible for individuals to collect global information in real social systems. Therefore,  this only
gives a macroscopic explanation of how a user's degree of activity and
an item's degree of popularity evolve. Moreover, the formation of cross links will
simultaneously affect the activity degree and  popularity degree.
This poses an interesting question: What is the microscopic mechanism
governing the growth of a user's activity degree and an item's popularity degree while giving rise to the various distribution observed?

There are two crucial questions to be considered. First, how are
the users activated to create new cross links? It is very
difficult to formulate the users' behaviors because human dynamics
are very complicated due to the inherent diversity of real world
circumstances. In the empirical analysis, the users with a larger degree of activity are more likely to create cross links. Moreover, it is found that the degree of activity is positively correlated to the social degrees \cite{activity}. In addition, the users' activities
are influenced by their neighbors' activities as the information can
spread along social links by user-to-user exchanges
\cite{socialinfluence1,socialinfluence,socialinfluence2,healthbehavior,socialinfluence3}. Therefore, we believe that the users with more friends are frequently activated as receiving more
information from neighbors, and the users interested in more items
are easily activated because they are very sensitive to stimuli.
Hence, we employ the random walk, starting from one user and
via either social or cross links, to select users, who will actively create cross links. The users with more friends and more items have a greater chance of being reached by random walk. It is found that random walk might be one possible micro mechanism governing the evolution of social networks\cite{microdynamics,measurePA}, and is equivalent to preferential attachment from a macro perspective \cite{measurePA,randomwalk}.

\begin{figure}[!ht]
\begin{center}
\includegraphics[width=5.5in]{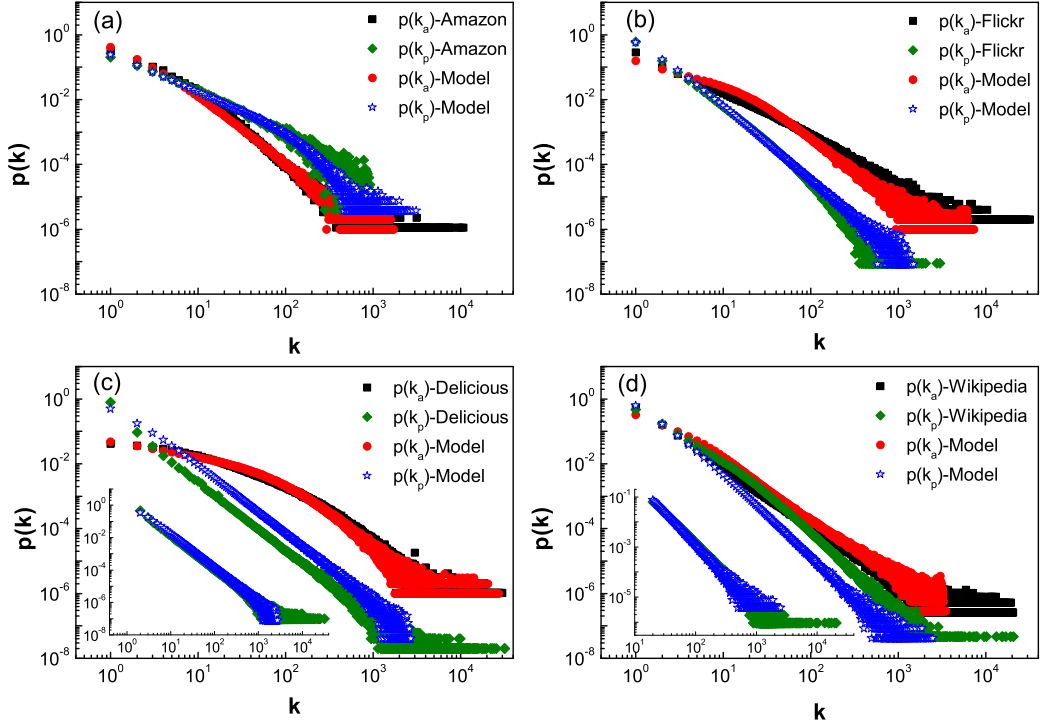}
\end{center}
\caption{{\bf Comparing the empirical distributions with those of the model.}  Distributions of activity degree $p(k_a)$ and popularity degree $p(k_p)$. (a) Amazon, where $m=5$, $n=7$ and $q=0.27$. (b) Flickr, where $m=30$, $n=30$ and $q=12$. (c) Delicious, where $m=5$, $n=120$ and
$q=25$. The inset is the distribution of degree of popularity for
$k_p\geq 2$. (d) Wikipedia, where $m=80$, $n=10$ and $q=7$. The inset is
the distribution of degree of popularity for $k_p\geq 11$.}\label{fig5}
\end{figure}

The second question to consider is: How do the activated users access the items? In Flickr, it has been found that over 80\% of new social links are formed between friends' friends and  over 50\% of new cross links are formed
between one user and his friends' favorite photos \cite{Flickr}. Moreover, the
probability of a user favorite-marking one photo increases with the
number of his friends who have favorite-marked the photo
\cite{Flickr}, indicating that the user is influenced by his friends
and reaches the photo via his friends. We therefore assume that the
users access the items via their friends by two-step random walk process.
It is also worth noting that the popular items are exposed to more users, so they have a higher probability of being reached by the random walk process than the unpopular items.

For simplicity, we made the following assumptions in our model: (1)
the users can befriend other users (see the example of Fig.
\ref{fig1}), (2) the activated users are selected by random walk
either via social or cross links, and (3) new links (except the
first links attached to newly added users and items) are formed
between one user and one of his second neighbors (either users or items).
In this way, the link growth process can be understand as two-step random
walks via the social links or via the cross links. One way to model this network is to select the activated users who will then actively create new links. Another is to select the target nodes (either users or items) that will
passively acquire new links. In our numerical simulation, we employ
a two-step random walk for simplicity.

Numerically, the topology evolves according to the following rules: (1) the initial network consists of a few users ($M_0$)
and items ($N_0$). The users form a small random social network,
while the items are randomly rated by the users. (2) At each time step, one new user is added into the system and randomly connected to one user and one of items rated by the user that is being connected to.
Meanwhile, $q$ new items are added into the system, each of which is
rated by one activated user selected by two-step random walk. (3)
At each time step, $m$ users are activated by using a
two-step random walk, and each of them connects to his second
neighbors by a two-step random walk via common friends or via common
items if they are not previously connected. (4) At each time step, $n$ users are activated by a two-step
random walk, and each of them connects to one of items rated by his
friends by a two-step random walk if he has not previously rated it.

\begin{figure}[!ht]
\begin{center}
\includegraphics[width=5.5in]{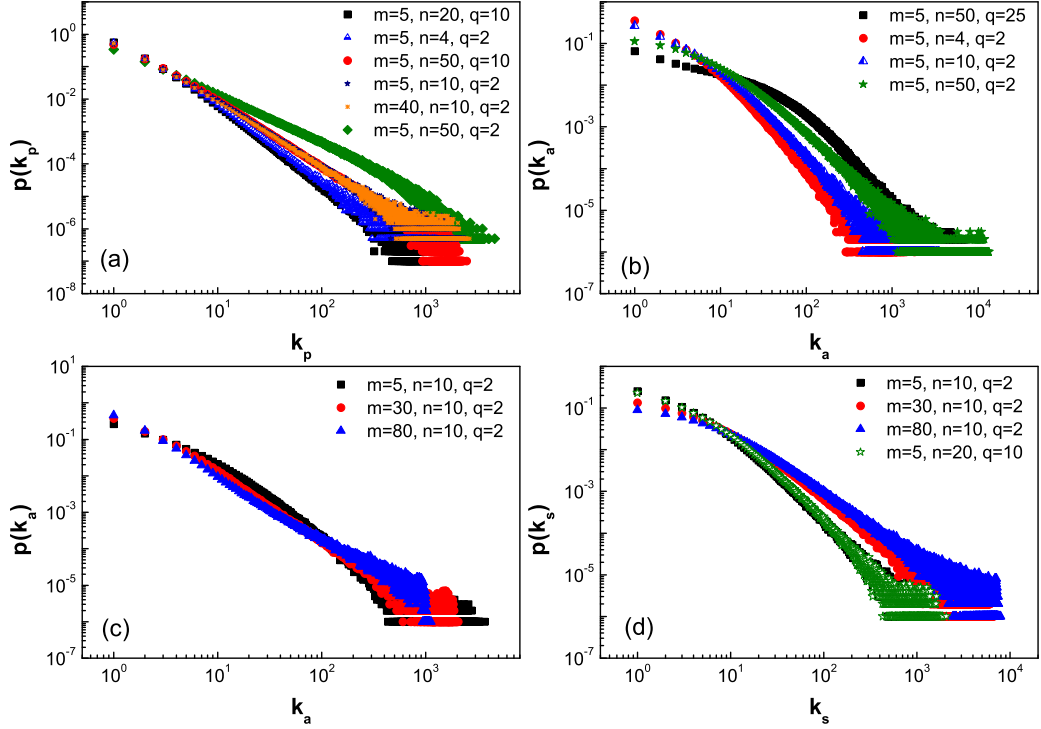}
\caption{{\bf Influence of parameters on the distribution of degree.} Influences of parameters
$n$ and $q$ on the distribution of popularity (a) as well as the
distribution of activity (b). Influences of parameter
$m$ on the distribution of activity (c) and the
distribution of social degree (d).}\label{fig6}
\end{center}
\end{figure}

We carried out numerical simulations to validate the model. We set
the parameters as follows: $M_0=100$ and $N_0=100$. We then ran the simulation
up to $M=100,000$. Based on the simple assumptions of random
walk, the model can reproduce the distributions of
activity degree $p(k_a)$ and popularity degree $p(k_p)$ observed in
the above four empirical networks. In Fig. \ref{fig5}, we compare the
degree distributions of the model with those of the empirical
networks. It is found that the distributions of the model networks are
qualitatively consistent with their counterparts of the empirical
networks. Though the distributions of popularity degree in the
model do not quantitatively match those of empirical networks as
shown in Figs. \ref{fig5} (c) and (d), the slops are consistent with
each other. In the insets of Figs. \ref{fig5} (c) and (d), we
compare the distributions of larger degree of popularity of the model
with those of the empirical networks. It is observed that they are approximately consistent with each other. These results indicate that our assumptions are reasonable that the users are activated by a two-step random walk, and subsequently
find items of interest to them in the same manner.

Figure \ref{fig6} displays the influences of the parameters on the
distributions. From Fig. \ref{fig6} (a), we can see that the ratio
between the number of new cross links $n$ and the number of new items
$q$ has an obvious influence on the distribution of popularity degree;
however, the distributions of activity degree  are greatly affected
by the parameter $n$ as seen in Fig. \ref{fig6} (b). Furthermore, the
distributions of activity degree depend on the number of new
social links $m$ to some extent, while the distributions of social
degree $p(k_s)$ are almost independent of the number of new cross
links $n$ and the number of new items $q$. In particular, the
distributions of social degree follow a power law, which is
qualitatively consistent with that of Flickr networks.

\section*{Discussion}
In this study, we first carried out an empirical analysis to four
empirical networks: Amazon, Flickr, Delicious and Wikipedia.
Our study revealed the growth patterns of the users' degrees of activity and
items' degrees of popularity within these networks, both of which follow the law of the rich-get-richer.
It was found that the users are likely to trace popular items, but the intensities are different for the users with different activity degrees. For example, the active users make a greater than average contributions to the unpopular items whereas the inactive users make a greater than average contributions to the popular items. Motivated by the empirical findings in
these four networks, we proposed an evolving model based on a two-step
random walk, which is able to qualitatively reproduce the activity
and popularity distributions observed in empirical networks. Based
on both the empirical analysis and the theoretical model, we believe that the
information spreading amongst individuals, which is simplified as
a two-step random walk in the model, could represent one possible micro mechanism governing the dynamic evolution of human activity and item popularity. Of course, the dynamics of human activity and item popularity are very complicated
due to the inherent diversity of human behaviors and the varying
nature of items. Hence, there may be other microscopic mechanisms
governing the dynamics of human activity and item popularity.

It should be noted that the results of our model are only qualitatively consistent with the empirical results. The
quantitative mismatch is due to the simplifications in our model.
For example, the users are only activated by a two-step random walk,
and then reach the items by another two-step random walk via friends.
In reality, the situation could be much more complicated. In
the empirical networks, for example, besides the stimulus of
neighbors, the occasional events can inspire users to
participate in network-relating events. For instance, the users can access photos through various other channels in Flickr, such as the list of interesting photos provided by the web site, the search engine, the links between
similar photos and so on. Furthermore, an item's particular attributes, such as being an award-winning picture, may disproportionally affect how quickly an item's popularity changes. These factors
have remarkable influence on the growth of human activity and item popularity. We
believe that if we consider more realistic factors in the model, we
can improve the performance of our model and obtain more helpful
insights in understanding the dynamics of human activity and item
popularity. These problems deserve further investigations in the
future.


\section*{Acknowledgments}
This work is supported by the Fundamental Research Funds for the Central Universities under Grant Nos. 2012RC0707 and 2012JBM067, National Natural Science Foundation of China under Grant Nos. 61174150, 71101009 and 61374175, NCET-09-0228 and the Doctoral Fund of the Ministry of Education
(20110003110027). 




\end{document}